\documentclass[aps,pra,twocolumn,showpacs,superscriptaddress,floatfix]{revtex4-1}
\usepackage[utf8]{inputenc} %useful to type directly diacritic characters
\usepackage{amssymb,amsmath}
\usepackage{graphicx}
\usepackage{siunitx}
\usepackage{xcolor}
\usepackage{physics}
\usepackage{hyperref} % Hyperlinks

\def\Rb87{^{87}\mathrm{Rb}}                             % Rb 87
\def\K40{^{40}\mathrm{K}}                    		    % K 40 

\usepackage[normalem]{ulem}

\begin{document}

%TC:ignore 

\title{Dynamical Structure Factor from Weak Measurements}

\author{E.~Altunta\c{s}}
\email{altuntas@ou.edu}
\affiliation{Center for Quantum Research and Technology, Homer L. Dodge Department of Physics and Astronomy, The University of Oklahoma, Norman, OK 73019, USA}
\affiliation{Joint Quantum Institute, National Institute of Standards and Technology, and University of Maryland, Gaithersburg, Maryland, 20899, USA}
\author{R.~G.~Lena}
\affiliation{Department of Physics and SUPA, University of Strathclyde, G4 0NG Glasgow, UK}
\author{S.~Flannigan }
\affiliation{Department of Physics and SUPA, University of Strathclyde, G4 0NG Glasgow, UK}
\author{A.~J.~Daley}
\email{andrew.daley@physics.ox.ac.uk}
\affiliation{Department of Physics and SUPA, University of Strathclyde, G4 0NG Glasgow, UK}
\affiliation{Department of Physics, University of Oxford, Clarendon Laboratory, OX1 3PU Oxford, UK}
\author{I.~B.~Spielman}
\email{ian.spielman@nist.gov}
\affiliation{Joint Quantum Institute, National Institute of Standards and Technology, and University of Maryland, Gaithersburg, Maryland, 20899, USA}
\date{\today}

\begin{abstract}
Much of our knowledge of quantum systems is encapsulated in the expectation value of Hermitian operators, experimentally obtained by averaging projective measurements.
However, dynamical properties are often described by products of operators evaluated at different times; such observables cannot be measured by individual projective measurements, which occur at a single time.
For example, the dynamical structure factor describes the propagation of density excitations, such as phonons, and is derived from the spatial density operator evaluated at different times.
In equilibrium systems this can be obtained by first exciting the system at a specific wavevector and frequency, then measuring the response.
Here, we describe an alternative approach using a pair of time-separated weak measurements, and analytically show that their cross-correlation function directly recovers the dynamical structure factor, for all systems, even far from equilibrium.
This general schema can be applied to obtain the cross-correlation function of any pair of weakly observable quantities.
We provide numerical confirmation of this technique with a matrix product states simulation of the one-dimensional Bose-Hubbard model, weakly measured by phase contrast imaging.
We explore the limits of the method and demonstrate its applicability to real experiments with limited imaging resolution.
\end{abstract}
%TC:endignore

\maketitle

%----------------------------------------------------------------------------------------
%	 Introduction
%----------------------------------------------------------------------------------------
 
The properties of many body quantum systems are often encoded in response functions that quantify the reaction of the system to weak external perturbations.
The dynamical structure factor (DSF) $S(q,\omega)$, describing a system's response to density perturbations with wavenumber $q$ and frequency $\omega$, derives from the expectation value $\langle\hat n(0,0)\hat n(x,t)\rangle$ of the spatial density operator $\hat n(x,t)$ at different times and positions~\cite{AshcroftBook1976}.
For equilibrium systems, such quantities are experimentally obtained by observing the system's response to a suitable perturbation.
For example neutron scattering gives access to $S(q,\omega)$ (possibly in a spin-dependent way) in materials~\cite{Sturm1993}, where it has contributed to the understanding of high temperature superconductivity~\cite{Hinkov2007}, and topological spin systems~\cite{Yao2018,Banerjee2018} to name a few.
Analogously for ultracold atoms, $S(q,\omega)$ can be obtained by Bragg scattering far-detuned laser light off atomic ensembles~\cite{Birkl1995}; this has shed light on weak- and strongly-interacting Bose-Einstein condensates~\cite{Steinhauer2002,Pino2011}, structural phase transitions~\cite{Landig2015}, and unitary Fermi gases~\cite{Biss2022}.
The spectral function, a related response function correlating fields rather than densities, can be measured using similar techniques~\cite{Bakr_2020, Bohrdt_2020, Bohrdt_2021}. 
By contrast, we focus on measuring density-density correlations $\langle\hat n(0,0)\hat n(x,t)\rangle$, and thereby $S(q,\omega)$, by weak quantum measurements alone~\cite{Xu2015}.

Unlike the methods described above, this technique is not limited to equilibrium systems, and is applicable to any state, pure or mixed, undergoing Hamiltonian dynamics.
This connects to previous work~\cite{Hung2011a}, that showed the static structure factor $S(q)$ can be obtained from the same-time density-density correlation function $\langle\hat n(0)\hat n(x)\rangle$ computed from simple projective measurements of density.
Because projective measurements collapse the wavefunction at a well-defined time, they prevent access to the two-time correlations we require. 

\begin{figure}[bt!]
\begin{center}
\includegraphics{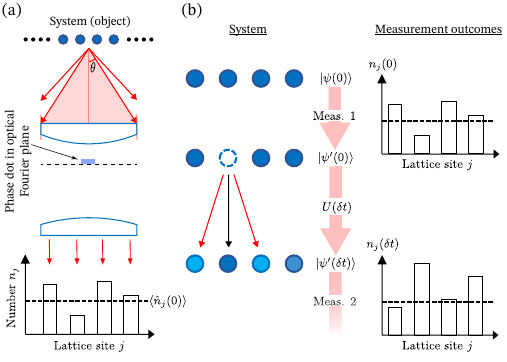}
\end{center}
\caption{System and measurement schema. 
(a) PCI as an example of homodyne detection. 
Probe light is first phase shifted by the atomic ensemble (light scattered by more than the acceptance angle $\theta$ is not detected); the un-scattered light is further phase shifted by a phase dot; and the resulting interference between the scattered and the un-scattered light is imaged.
The detected signal (histogram) has contributions from the operator expectation value (dashed line) as well as projection noise.
(b) A pair of time-separated weak measurements.
The first measurement creates excitations that are correlated with the measurement outcome and propagate for a time $\delta t$ prior to a second measurement.
}
\label{Fig1:Measurement_Schematic}
\end{figure}

Generalized quantum measurements, however, allow for weak measurements that minimally disturb the system, but in exchange provide only limited information. 
These have long been associated with continuous monitoring of both open and closed quantum systems~\cite{Guerlin2007,Minev2019}, and are essential for closed loop quantum control~\cite{Wiseman2009}.
Here, we make use of weak measurements in a many-body context, where they will allow the extraction of multi-time correlation functions, valid independent of quantum statistics and dimensionality.
We specifically focus on homodyne weak measurement schemes [for example, realized by phase-contrast imaging (PCI) as shown in Fig.~\ref{Fig1:Measurement_Schematic}(a)] of atoms in a 1D lattice that report $n_j(t)$, a noisy estimate of number at lattice site $j$ density, and yield post-measurement wavefunctions that incorporate density fluctuations inferred from the measurement.
The excitations resulting from the first weak measurement at time $t=0$ propagate for a time delay $\delta t$ prior to a second weak measurement of density giving $n_{j^\prime}(\delta t)$, as in Fig.~\ref{Fig1:Measurement_Schematic}(b). 
Intuitively the second measurement outcome includes a contribution from the system's response to the first measurement, and we analytically confirm that the ensemble averaged~\footnote{The ensemble can include trajectories for a single initial pure state, or different initial states sampled from a specified distribution.} cross-correlations between these observations $\overline{n_j(0)n_{j^\prime}(\delta t)}$, converges to the real part of the un-equal time expectation value: ${\rm Re}\left[\langle \hat n_j(0) \hat n_{j^\prime}(\delta t)\rangle\right]$.
This interpretation is similar to that presented Ref.~\onlinecite{Xu2015} in the context of photodetection theory.

We numerically study an implementation of this approach for the 1D Bose-Hubbard (BH) model, and show that in the limit of vanishing measurement strength $S(q,\omega)$ obtained by the weak measurement process converges to that directly computed from the underlying wavefunctions.
We conclude by showing that this method robustly captures long wavelength excitations even with realistic imaging limitations included.

%----------------------------------------------------------------------------------------
%	 Homodyne detection
%----------------------------------------------------------------------------------------

{\it Homodyne detection}--
Our weak measurement protocol is based on homodyne detection, which could be experimentally implemented using PCI as schematically depicted in Fig.~\ref{Fig1:Measurement_Schematic}(a).
When illuminated with coherent light of wavelength $\lambda$, an atomic ensemble in the object plane diffracts part of the probe laser by imprinting a position-dependent phase shift onto it.
PCI is usually operated in the far-detuned limit, where absorption can be ignored, so the intensity of the probe laser just after having traversed the atomic ensemble is unchanged.
By design, PCI is sensitive to changes in the phase quadrature: a phase dot at the Fourier plane of a Keplerian telescope phase shifts the unscattered light by a known phase value (typically $\pi/2$) while leaving the scattered light unchanged~\cite{Altuntas2021}. 
PCI is an interferometric method in which the unscattered probe serves as the reference beam (i.e., local oscillator), while the scattered component carries information about the atomic ensemble.
In the image plane, the interference between scattered light and the phase-shifted probe changes the detected intensity by an amount proportional to the atomic density.
The usual ``photon shot noise'' in this image is the projection noise associated with detecting the intensity of a coherent state.
For a bright beam, this leads to backaction described by a Kraus operator that is a Gaussian function of the atomic density operator.

%----------------------------------------------------------------------------------------
%	Fig1
%----------------------------------------------------------------------------------------

%----------------------------------------------------------------------------------------
%	 Analytical derivation of structure factor
%----------------------------------------------------------------------------------------

{\it Analytical derivation}---To be concrete, we consider bosonic atoms in a 1D lattice with an arbitrary native system Hamiltonian $\hat{H}$, responsible for coherent dynamics.
These dynamics are interjected by brief homodyne measurements of the atomic density, with durations short compared to the time scale for coherent evolution.
Each measurement, at a discrete time $t$, gives an outcome 
\begin{align}
n_{j,t} &= \langle \hat n_j(t)\rangle + \frac{m_{j,t}}{2\Gamma^{1/2}}, \label{eq:measure}
\end{align}
where $\Gamma$ is the measurement strength; $\hat n_j$ is the number operator at position $j$; and $\langle \hat n_j(t)\rangle \equiv \bra{\psi(t)} \hat n_j\ket{\psi(t)}$ is the Schrödinger picture expectation value of number.
The random variable $m_{j,t}$, with variance $\overline{m_{j_1,t_1} m_{j_2,t_2}} = \delta_{j_1,j_2}\delta_{t_1,t_2}$, describes spatially and temporally uncorrelated quantum projection noise.
For small $\Gamma$, the state conditioned on this measurement outcome~\cite{Plenio1998} is 
\begin{align}
\ket{\psi^\prime(t)}\!=&\!\bigg(\!1 + \Gamma^{1/2}\sum_j \delta \hat n_{j,t} m_{j,t} - \frac{\Gamma}{2}\sum_j \delta \hat n_{j,t}^2\!\bigg)\!\ket{\psi(t)} \label{eq:hom_det}
\end{align}
in terms of the difference operator ${\delta \hat n_{j,t}} \equiv \hat n_j - \langle \hat n_j(t)\rangle$. 
In following expressions we adopt a notation where expectation values with a prime
\begin{align} 
    \langle \hat{n}_{j^\prime}(\delta t) \rangle^\prime &= \bra{ \psi^\prime(0)} \hat U^\dagger(\delta t) \hat{n}_{j^\prime} \hat U(\delta t) \ket{\psi^\prime(0)}
    \label{eq:PostMeasure}
\end{align}
are with respect to the post measurement initial state.

In our protocol [illustrated in Fig.~\ref{Fig1:Measurement_Schematic}(b)], we (1) perform an initial measurement at time $t=0$; (2) allow the system to undergo unitary evolution described by $\hat U(\delta t)=\exp\big(-i \hat H \delta t / \hbar\big)$ for a time $\delta t$; and (3) perform a second measurement.
We focus on the cross-correlation of the measurement results
\begin{align}\label{eq:corr_sigs}
    \overline{n_{j,0} n_{j^\prime,\delta t}} &= \frac{1}{2\Gamma^{1/2}} \overline{m_{j,0} \langle \hat{n}_{j^\prime}(\delta t) \rangle^\prime} + \overline{\langle \hat{n}_{j}(0) \rangle \langle \hat{n}_{j^\prime}(\delta t) \rangle^\prime }.
\end{align}
Terms such as $\overline{m_{j^\prime, \delta t} \langle \hat n_{j} (0) \rangle }$ are absent in this expression because the number at earlier times is uncorrelated with measurement noise at later times.
By contrast, $\langle \hat{n}_{j^\prime}(\delta t) \rangle$ can be correlated with the noise $m_{j,0}$.
Both terms in Eq.~\eqref{eq:corr_sigs} have clear physical meaning: the first term identifies excitations created by measurement backaction, while the second term is simply the correlation between the initial density and the averaged final density that one might naively expect.

To lowest order in $\Gamma$, the expectation value of the number at time $\delta t$ is
\begin{align}
    \langle \hat{n}_{j^\prime }(\delta t) \rangle^\prime &= \langle \hat{n}_{j^\prime}(\delta t) \rangle +  \Gamma^{1/2}\sum_{j}m_{j,0} \big[ \langle \hat{n}_{j}(0) \hat{n}_{j^\prime}(\delta t) \rangle \nonumber \\
    &\ \ \ + \langle \hat{n}_{j^\prime}(\delta t) \hat{n}_{j}(0) \rangle - 2  \langle \hat{n}_{j}(0)\rangle \langle \hat{n}_{j^\prime}(\delta t)  \rangle \big]. \label{eq:measured_density}
\end{align}
Substituting this expression into Eq.~\eqref{eq:corr_sigs} yields $\overline{n_{j,0} n_{j^\prime,\delta t}} = {\rm Re}\left[ \langle \hat{n}_j(0) \hat{n}_{j^\prime}(\delta t) \rangle \right]$, equal to the expectation value of the Hermitian part of $\hat{n}_j(0) \hat{n}_{j^\prime}(\delta t)$. 
This gives direct access to the spatially averaged correlation function
\begin{align} \label{eq:relat}
G_{\delta j}(\delta t) &= \frac{1}{N} \sum_j {\rm Re}\left[ \langle \hat{n}_j(0) \hat{n}_{j + \delta j}(\delta t) \rangle \right],
\end{align}
as a function of spatial displacement $\delta j$, known as the Van Hove function~\cite{Van-Hove1954} in x-ray and neutron scattering.
This demonstrates our central observation: for $\Gamma\ll 1$, correlating subsequent measurement results yields time-separated correlation functions (the appendix comments on the consequence of stronger measurements.).
This is a generic observation valid for pairwise combinations of weak measurements  (potentially of different observables) with the general structure in Eqs.~\eqref{eq:measure}-\eqref{eq:hom_det}. 

The DSF
\begin{align}
    S(q, \omega) &= \sum_{\delta j} \int_0^{\infty} G_{\delta j}(\delta t) e^{i(\omega \delta t - q \delta j)} d \delta t \label{eq:DSF_from_measurements}
\end{align}
is the Fourier transform of the correlation function. 
In the subsequent sections, we present concrete numerical examples as to how accurately this quantity can be captured with a realistic measurement strength.

%----------------------------------------------------------------------------------------
%	 1D Hubbard chain
%----------------------------------------------------------------------------------------

%----------------------------------------------------------------------------------------
%	Fig2
%----------------------------------------------------------------------------------------

\begin{figure}[t]
\includegraphics{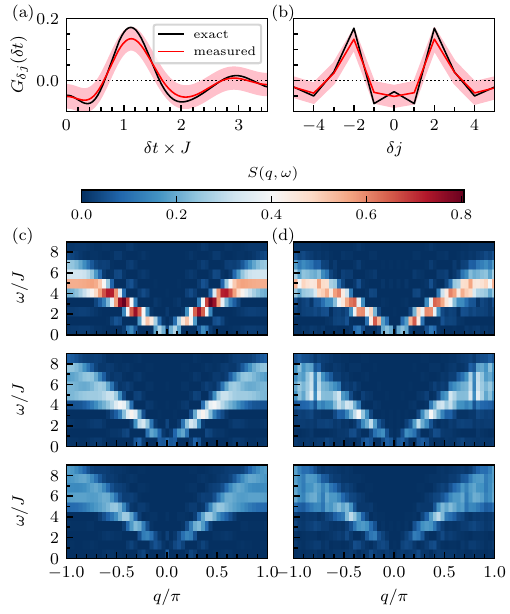}
\caption{Averaged correlation function and DSF of a 51-site 1D BH chain with measurement strength $\Gamma = 0.1$, $50$ trajectories, and the noise contribution removed from the second measurement.
Statistical uncertainties and spatial resolution limits are detailed in the text and presented in Figs.~\ref{Fig_Error_Comp} and \ref{Fig_NA_Results} respectively.
(a,b) Correlation signal with $U/J=2$ at: (a) fixed $\delta j = 2$, and (b) at fixed $\delta t \times J = 1.17$. 
The black curves were computed directly from MPS wavefunctions, while the red curves employed Eq.~\eqref{eq:relat} (red bands reflect the single-$\sigma$ uncertainties derived from our 50 trajectories).
(c, d) DSF computed for $U/J=2$, $4$, and $5$ (top to bottom); (c) DSF from MPS wavefunctions, and (d) DSF obtained from Eq.~\eqref{eq:DSF_from_measurements}.
}
\label{Fig_BH_Results}
\end{figure}

{\it Weakly measured BH system}---We now turn to the BH model with Hamiltonian
\begin{align*}
\hat{H}_{\rm BH} =& -\!J \sum_j \left( \hat{a}^{\dagger}_j \hat{a}_{j+1} + {\rm H.c.} \right) + \frac{U}{2}\sum_j \hat{n}_j \left( \hat{n}_j + 1 \right), %\label{BH_MOD}
\end{align*}
where $\hat{a}^{\dagger}_j$ describes the creation of a boson on site $m$ and $\hat{n}_j = \hat{a}^{\dagger}_j \hat{a}_j$ is the associated number operator.
We limit the numerical study to 1D and simulate the dynamics using the time-dependent variational principle applied to matrix product states (MPS)~\cite{Bridgeman2017}.

We numerically obtain the ground state $\ket{\psi(0)}$, and then initiate a specific trajectory by stochastically selecting a realization of the projection noise $m_{j,t=0}$. This allows us to compute the associated post-measurement state $\ket{\psi^\prime(0)}$ using Eq.~\eqref{eq:hom_det} and measurement outcome $n_{j,0}$.
The next step is to evolve the resulting state according to $\hat{H}_{\rm BH}$ for a time $\delta t$, obtain $\langle\hat n_{j}(\delta t)\rangle$, select a second noise realization $m_{j,\delta t}$, and thereby obtain the second measurement outcome $n_{j, \delta t}$.
We then average the correlation signal [Eq.~\eqref{eq:corr_sigs}] over trajectories. 
To increase sampling efficiency in the numerical simulations, we use each initial noise realization $m_{j,t=0}$ to generate the full collection of second measurement outcomes, by computing $\langle\hat n_{j}(\delta t)\rangle$ and $m_{j,\delta t}$ at every desired time increment. 
When reporting correlation functions, we further enhance the sampling efficiency by omitting the noise contribution to the second measurement, giving $n_{j,\delta t} = \langle \hat n_j(\delta t)\rangle$.
The number of trajectories quoted in the figure captions reflects the number of distinct initial noise realizations. 

Figure~\ref{Fig_BH_Results} plots the results of these simulations (see caption for simulation parameters); (a) and (b) show $\delta t$ and $\delta j$ cross-sections of the correlation function $G_{\delta j}(\delta t)$ with quantities extracted directly from MPS wavefunctions (black), and those obtained using Eq.~\eqref{eq:relat} (red) agreeing within the statistical uncertainty (error bands).
The duration of the MPS simulations are limited to the short times--- $\delta t \times  J \lesssim 3$ for current parameters---for which they are numerically exact (i.e., converged in truncation error per-timestep); this blocks numerical access to small $\omega$ when computing $S(q,\omega)$.
Longer times (and therefore smaller $\omega$) are experimentally accessible, however, finite size effects will appear for large $\delta t$.
For example, at $U/J = 2$ the correlation signal nominally spreads by approximately two lattice sites per time increment of $\delta t \times J $. 
Then, in a 51-site 1D BH chain, an excitation initially located at the center of the system will reach the boundary in $\approx 12$ tunneling times.

Panels (c) and (d) confirm that $S(q,\omega)$ extracted from the MPS wavefunction (left) is in correspondence with that obtained from the correlation signal (right).
The key features of the excitation spectrum~\cite{Fisher1989} are present: in the superfluid phase ($U/J=2$) a gapless phonon mode (linear for small $\omega$) is present; and in the Mott insulating phase ($U/J=4, 5$) a energy gap clearly appears.
In both cases the DSF contains sharp spectra features derived from well defined quasiparticle  excitations: phonons in the superfluid phase giving way to particle-like excitations in the Mott phase.
The artifacts present in the correlations signal in Fig.~\ref{Fig_BH_Results}(d) arise from a hybrid term that combines both statistical and systematic uncertainties (see Appendix A.).

%----------------------------------------------------------------------------------------
%	Systematic errors
%----------------------------------------------------------------------------------------

\begin{figure}[bt]
\includegraphics{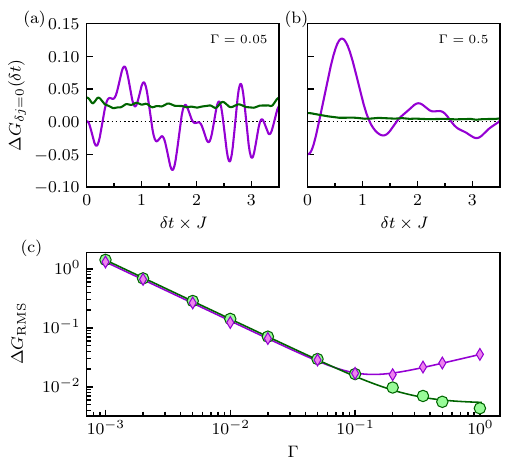}
\caption{
Statistical and systematic uncertainties using 50 trajectories and $U/J = 2$.
(a,b) Comparison of statistical uncertainty (green) and total error (purple) in $\Delta G_{\delta j=0}(\delta t)$ for $\Gamma = 0.05$ and $\Gamma = 0.5$.
(c) Statistical (green) and total (purple) uncertainties averaged over $\delta t$ and $\delta j$ along with fits (curves) to the scaling behavior described in the text.
The statistical uncertainties associated with these averages are smaller than the plotted markers.
}  
\label{Fig_Error_Comp}
\end{figure}

{\it Systematic and statistical uncertainties}---Any determination of $S(q,\omega)$ will have statistical uncertainties resulting from employing only a finite number of trajectories given the noise inherent in the measurement process.
In addition, our expressions for spatial correlations and the DSF are only strictly valid when $\Gamma\rightarrow0$, giving corrections to  Eq.~\eqref{eq:corr_sigs} at order $\Gamma^{1/2}$, i.e., systematic uncertainties.
Both measurements [Eq.~\eqref{eq:measure}] have statistically independent noise contributions $\propto \Gamma^{-1/2}$. (While noise on the second measurement can be neglected when computing averaged quantities, it must be included when computing uncertainties.)
Together these contribute a $\propto\Gamma^{-1}$ statistical uncertainty to the correlation function $G_{\delta j}(\delta t)$.
In the strong-measurement limit the noise contribution of the second measurement can be neglected, and the remaining statistical uncertainty becomes independent of $\Gamma$.

Figure~\ref{Fig_Error_Comp}(a) and (b) quantify the statistical and total uncertainties at small and large $\Gamma$ respectively. 
The green curves plot the standard error of the mean across 50 trajectories (i.e., statistical uncertainty), and the purple curves plot the difference between $G_{\delta j}(\delta t)$ determined exactly from MPS wavefunctions and from our weak measurement approach (with contributions from both statistical and systematic sources).
At small $\Gamma$, statistical uncertainties exceed systematic effects, making the total and statistical uncertainties comparable; for large $\Gamma$ the relative importance reverses, so the total uncertainty exceeds the statistical contribution. 
This is summarized in (c) which plots the overall uncertainties $\Delta G_{\rm RMS}$, the quadratic mean averaged across all $\delta t$ and the spatial range $\delta j = -10$ to $10$, where there is significant signal.
As expected, the statistical uncertainty (green) initially decreases as $\Gamma^{-1}$ before beginning to saturate at large $\Gamma$.
As suggested by (a) and (b), the combined uncertainty (purple) is dominated by statistical effects for small $\Gamma$ and systematic effects for large $\Gamma$, where it scales as $\propto \Gamma^{1/2}$.
The solid curves in Fig.~\ref{Fig_Error_Comp}(c) confirm these observations with fits to the quadratic mean of the asymptotic scaling behaviors: $\big[\left(A \Gamma^{-1}\right)^2 + B^2\big]^{1/2}$ and $\big[\left(A \Gamma^{-1}\right)^2 + \left(C \Gamma^{1/2}\right)^2\big]^{1/2}$ for statistical and total uncertainties respectively (with fit parameters $A, B$, and $C$).
In practice an optimal measurement strategy balances these two sources of uncertainty; in our 50-trajectory simulations these contributions are comparable at $\Gamma = 0.1$.

{\it Resolution limits}---
Even aside from technical considerations associated with imperfect imaging system design, realistic imaging systems only capture light scattered at angles below an angular acceptance $\theta$; this introduces a momentum-space cutoff $k_{\rm max} = 2\pi\sin\theta/\lambda$~\footnote{In addition, for systems that are spatially extended along the imaging axis by more than the depth of field, the whole system cannot be simultaneously in focus; this leads to an effective reduction in the NA.  Here our focus is on very strong transverse confinement, making this second effect negligible.}.
As a result, far-field imaging such as PCI provides no information for wavenumbers above $k_{\rm max}$, and for this reason the current generation of PCI experiments cannot perfectly resolve individual lattice sites.

%----------------------------------------------------------------------------------------
%	Fig4
%----------------------------------------------------------------------------------------
\begin{figure}[bt]
\includegraphics{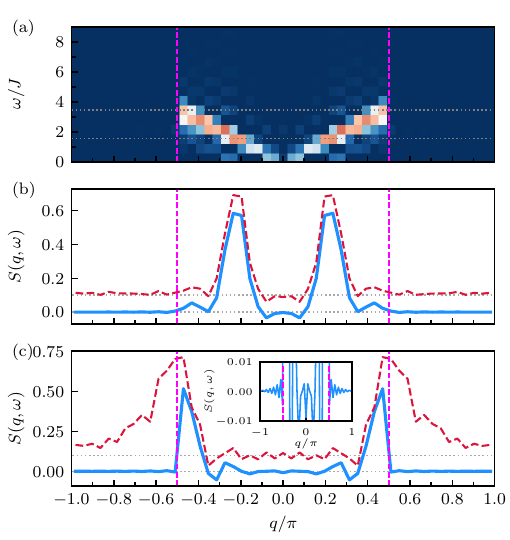}
\caption{
DSF derived from measurements with imaging resolution effects ($k_{\rm max} = 0.5$, vertical dashed lines) computed for $U/J =2$.
As in Fig.~\ref{Fig_BH_Results}, the noise contribution is removed from the second measurement.
The resolution-limited DSF is plotted in (a) using the same color scale as in Fig.~\ref{Fig_BH_Results}, while (b) and (c) show cross sections taken at $\omega/J = 1.57$ and $\omega/J = 3.46$ and  with (blue) and without (red, offset for clarity) resolution limits.
The inset in (c) is vertically expanded to show artifacts outside $k_{\rm max}$.
}
\label{Fig_NA_Results}
\end{figure}

Given the inevitability of such resolution limits, we numerically simulated the impact of finite resolution by applying Fourier cutoffs (removing all Fourier components beyond some $k_{\rm max}$) to the measurement outcomes $n_{j,t}$ prior to computing correlation functions.
Figure~\ref{Fig_NA_Results}(a) shows the DSF obtained in this way is largely unchanged for $|k| < k_{\rm max}$ (vertical magenta dashed lines) but is abruptly cutoff at $k_{\rm max}$. 
Panels (b) and (c) continue by plotting cross sections taken at  $\omega/J = 1.57$ and $\omega/J = 3.46$ [horizontal dashed lines in(a)] both with (blue) and without (red) this cutoff.
In (b) the sharp spectral feature is clearly resolved and the DSF is essentially unchanged for $|k| < k_{\rm max}$.
In (c) the cutoff intersects the spectral feature and additional deviations appear: for smaller $k$ the DSF differs from the no-cutoff case, and the $|k| > k_{\rm max}$ signal (inset) oscillates around zero.
Taken together these data confirm that experimentally realistic imaging systems can obtain the DSF associated with long-wavelength excitations up to their $k_{\rm max}$, with worse-case artifacts at the $\approx5\ \%$ level.

%----------------------------------------------------------------------------------------
%	 Discussion and Outlook
%----------------------------------------------------------------------------------------

{\it Conclusion and outlook}---Here, we described an experimentally realistic technique employing a pair of weak measurements to extract $S(q,\omega)$, a four-field correlation function; although our study focused on degenerate Bose gases, this method is equally applicable to Fermi gases, Bose-Fermi mixtures, and spin systems.
Other techniques for measuring $S(q,\omega)$, such as Bragg spectroscopy, detect the system's response to experimentally induced perturbations.
Therefore, separate technical infrastructure is required to create excitations at each desired $q$ and $\omega$, and then to measure the response.
From this practical perspective, our method reduces experimental complexity as it relies on measurements alone. 
We note that systems described by non-Hermitian evolution can also provide access to correlation functions of this type~\cite{Schuckert2020,Pan2020}.

As briefly noted above, this approach can be be used to obtain two-time correlation functions of any pair of Hermitian operators that can be weakly measured.
Such a process yields the expectation value of the $\mathrm{O}(\Gamma^{1/2})$ backaction operator of the first measurement [e.g., $\delta {\hat n}_j$ in Eq.~\eqref{eq:hom_det}] multiplied by the measurement operator of the second measurement [e.g, $\hat n_{j^\prime}$ in Eq.~\eqref{eq:PostMeasure}].
As a straightforward example, in spinor systems quantities such the ``spin Van Hove function'' $\langle(\hat n_{\uparrow,j,0} - \hat n_{\downarrow,j,0})(\hat n_{\uparrow,j',\delta t} - \hat n_{\downarrow,j',\delta t})\rangle$ (essentially correlations in the $\hat \sigma_z$ magnetization) could be easily measured by performing PCI with the probe laser red-detuned with respect to one spin state and blue-detuned with respect to the other~\cite{Shin2006a}.
Measuring $\sigma_x$ and $\sigma_y$ magnetization requires additional coherent control to transform the desired quantities to and from the measurement basis.
Other examples include spin-spin cross-correlation functions such as $\langle \hat n_{j, \uparrow} \hat n_{j^\prime,\downarrow}\rangle$ in multi-component (or even multi-species) systems and even hybrid spatial-momentum correlations $\langle \hat n_{j} \hat n_{k}\rangle$ that combine the momentum space number density $\hat n_k$ with the real space density. 

A straightforward extension of this scheme to three measurements provides access to a weak-measurement Leggett-Garg~\cite{Leggett1985,Jordan2006} correlation function $B_{j_1, j_2} = \overline{n_{j_1, t_1} n_{j_2,t_2}} + \overline{n_{j_1, t_2} n_{j_2,t_3}} - \overline{n_{j_1, t_1} n_{j_2,t_3}}$.
Leggett-Garg correlation functions obey Bells inequality-like relations that distinguish between classical and quantum correlations.
Here violations of a Leggett-Garg inequality may identify non-trivial entanglement in self-equilibrating closed systems, even when the eigenstate thermalization hypothesis is otherwise valid for local degrees of freedom~\cite{Rigol2008}.

Extensions of this approach that directly correlate more than two measurements access higher order correlation functions, however, the resulting signal derives from higher order moments of the random variable $m_{j,t}$ and are dwarfed by two-point correlators (with sufficient statistics higher same-time correlations have been observed~\cite{Schweigler2017}).
It is possible, therefore, that such extensions may even yield out of time ordered correlations with their ability to quantify quantum chaos and entanglement spreading~\cite{Rozenbaum2017}.

%----------------------------------------------------------------------------------------
%	 Acknowledgements
%----------------------------------------------------------------------------------------
%TC:ignore
\section*{Acknowledgments}
The authors thank J.~K.~Thompson for productive discussions, D. Barker and L. P. Garcia-Pintos for carefully reading the manuscript.
Additionally, L. P. Garcia-Pintos identified the connection to Leggett-Garg inequalities.
This work was partially supported by the National Institute of Standards and Technology; the National Science Foundation through the Quantum Leap Challenge Institute for Robust Quantum Simulation (grant OMA-2120757); and the Air Force Office of Scientific Research Multidisciplinary University Research Initiative ``RAPSYDY in Q'' (FA9550-22-1-0339).

\section*{Disclosures}
The authors declare no conflicts of interest.

\section*{Data availability statement} 
The data that support the findings of this study are available upon reasonable request from the authors. 

\onecolumngrid 
\appendix

\section{Increased measurement strength}

In the main body of the text, we introduced a measurement model [Eqs.~\eqref{eq:measure}-\eqref{eq:hom_det}] that is valid up to second order in $\Gamma^{1/2}$.
This model describes both measurement outcomes and the associated conditional change in the system's wavefunction.
We then related the Van Hove function $G_{\delta j}(t')$ to the cross-correlation function of measurement outcomes at order $\mathrm{O}(\Gamma^{1/2})$ in the wavefunction update rule, and performed an ensemble average in the limit of infinitely many trajectories.
In this appendix, we quantify the next-order contributions to $G_{\delta j}(t')$, and examine the impact of employing a finite number $M$ of trajectories.

\subsection{Model}

In our initial exposition, we considered correlation functions such as $\overline{n_{j,0} n_{j',t'}}$, for which the logic of the argument was most transparent.
Here, we instead focus on noise correlations, i.e., $\overline{\delta n_{j,0} \delta n_{j',t'}}$ with $\delta n_{j,t} \equiv n_{j,t} - \overline{n_{j,t}}$; this choice leaves the underlying reasoning and conclusions unchanged but leads to greatly simplified expressions.

In terms of noise variables, the measurement model becomes
\begin{align}
\delta n_{j,t} &= \left(\langle \hat n_{j,t}\rangle - \overline{\langle \hat n_{j,t}\rangle}\right)+ \frac{m_{j,t}}{2 \Gamma^{1/2}}, & {\rm and} && \ket{\psi'(t)} &= \left(1 + \Gamma^{1/2}\sum_{j} m_{j,t} \delta \hat n_{j,t} - \frac{\Gamma}{2} \sum_{j} \delta \hat n_{j,t}^2 \right)\ket{\psi(t)} \label{eq:app:model}
\end{align}
for the weak measurement of atom number $\hat{n}_{j,t}$ at site $j$ and time $t$ with measurement strength $\Gamma \ll 1$.
Measurement noise is described by the random variable $m_{j,t}$, which has zero mean and covariance $\overline{m_{j,t} m_{j',t'}} = \delta_{j,j'}\delta_{t,t'}$.
Meanwhile, quantum fluctuations about $\langle \hat{n}_{j,t}\rangle$ are described by the operator $\delta\hat{n}_{j,t} \equiv \hat{n}_{j,t} - \langle \hat{n}_{j,t}\rangle$, i.e.,  $\langle \delta\hat{n}_{j,t} \rangle = 0$.
These expressions show both how the weak measurement result $n_{j,t}$ differs from the operator expectation value $\langle \hat{n}_{j,t}\rangle$ and how $\ket{\psi(t)}$ is conditionally updated to $\ket{\psi'(t)}$ based on the measurement outcome.

In general, the existence of time subscript on operators, such as $\hat{n}_{j}$ versus $\hat{n}_{j,t}$, is used to distinguish between Schrödinger and Heisenberg picture operators.
This convention fails for $\delta\hat{n}_{j,t}$ which explicitly depends on $t$ via $\langle \hat{n}_{j,t}\rangle$ independent of the picture; for example in Eq.~\eqref{eq:app:model} operators are in the Schrödinger picture.
(Note: as compared to the notation in the main body of the text, here time appears via a subscript on operators, i.e., $\hat{n}_{j}(t) \rightarrow \hat{n}_{j,t}$; this is to allow expressions such as in Eq.~\eqref{eq:app:density} to fit on a single line.)

The density-density correlation function requires knowledge of $\langle \hat n_{j,0}\rangle$ as well as $\langle \hat n_{j',t'}\rangle'$, which is influenced by measurement backaction and then by unitary evolution.
To order $\mathrm{O}(\Gamma)$, this latter quantity is
%Following the main text notation, this would be \$langle \hat{n}_{j^\prime }(\delta t) \rangle^\prime = langle \hat{n}_{j^\prime }(t) \rangle^\prime$. See Equ.~\eqref{eq:PostMeasure}}

\begin{align}
\langle \hat n_{j',t'}\rangle' &= \bra{\psi(0)}\left[1 + \Gamma^{1/2}\sum_{j_1} m_{j_1,0} \delta \hat n_{j_1,0} - \frac{\Gamma}{2} \sum_{j_1} \delta \hat n_{j_1,0}^2\right] \hat n_{j',t'}\left[ 1 + \Gamma^{1/2}\sum_{j_2} m_{j_2,0} \delta \hat n_{j_2,0} - \frac{\Gamma}{2} \sum_{j_2} \delta \hat n_{j_2,0}^2\right]\ket{\psi(0)} \nonumber \\
&= \langle \hat n_{j',t'} \rangle + 2 \Gamma^{1/2} \sum_{j_1} m_{j_1,0} {\rm Re}\left( \langle \delta \hat n_{j_1,0} \hat n_{j',t'} \rangle \right) - \Gamma\sum_{j_1} {\rm Re}\left(\langle \delta \hat n_{j_1,0}^2 \hat n_{j',t'}\rangle \right) + \Gamma \sum_{j_1,j_2} m_{j_1,0} m_{j_2,0} \langle\delta \hat n_{j_1,0} \hat n_{j',t'} \delta \hat n_{j_2,0}\rangle\nonumber\\
&\approx \langle \hat n_{j',t'} \rangle + 2 \Gamma^{1/2} \sum_{j_1} m_{j_1,0} {\rm Re}\left( \langle \delta \hat n_{j_1,0} \hat n_{j',t'} \rangle \right) - \Gamma\underbrace{\sum_{j_1} {\rm Re}\left(\langle \delta \hat n_{j_1,0}^2 \hat n_{j',t'}\rangle  - \langle\delta \hat n_{j_1,0} \hat n_{j',t'} \delta \hat n_{j_1,0}\rangle \right)}_{\equiv L_{j',t'}},\label{eq:app:density}
\end{align}
where we defined the $\mathrm{O}(\Gamma)$ term as $L_{j',t'}$ owing to its formal similarity to the Lindblad term in master equations.  
In simplifying $L_{j',t'}$, we replaced $m_{j_1,0} m_{j_2,0}\rightarrow \delta_{j_1,j_2}$ as suggested by Itô calculus  [a similar replacement was used when deriving the wavefunction update rule's $\mathrm{O}(\Gamma)$ term].
In analogy with the wavefunction update rule, only the $\mathrm{O}(\Gamma^{1/2})$ term depends on the random variable, leading to the simple expression 
\begin{align}
\langle \hat n_{j',t'}\rangle' - \overline{\langle \hat n_{j',t'}\rangle'} &= 2 \Gamma^{1/2} \sum_{j_1} m_{j_1,0} {\rm Re}\left( \langle \delta \hat n_{j_1,0} \delta \hat n_{j',t'} \rangle \right)\label{eq:app:evolved_noise_operator}
\end{align}
for fluctuations, thereby recovering Eq.~\eqref{eq:measured_density}.

\subsection{Statistical uncertainties}

With expressions for $\delta n_{j,0}$ and $\delta n_{j',t'}$ in hand, we now examine the cross correlation function
\begin{align}
\delta n_{j,0} \delta n_{j',t'} & = \frac{m_{j,0}}{2 \Gamma^{1/2}}\left[\left(\langle \hat n_{j',t'}\rangle' - \overline{\langle \hat n_{j',t'}\rangle'}\right) + \frac{m_{j',t'}}{2 \Gamma^{1/2}}\right] + \mathrm{O}(\Gamma^{1/2}) \nonumber\\
&= \frac{1}{4 \Gamma} m_{j,0} m_{j',t'} + \sum_{j_1}  m_{j,0} m_{j_1,0} {\rm Re}\left( \langle \delta \hat n_{j_1,0} \delta \hat n_{j',t'} \rangle \right).
\end{align}
The absence of any systematic artifacts in this un-averaged CCF through order $\mathrm{O}(\Gamma^{1/2})$ results from the complete cancellation of the $\mathrm{O}(\Gamma)$ term in Eq.~\eqref{eq:app:evolved_noise_operator}.

Performing the ensemble average then yields the mean
\begin{align}
\overline{\delta n_{j,0} \delta n_{j',t'}} &= {\rm Re}\left(\langle \delta \hat n_{j,0} \delta \hat n_{j',t'} \rangle\right),
\end{align}
and the noise-variance
\begin{align}
{\rm var} = \overline{(\delta n_{j,0} \delta n_{j',t'})^2} - \overline{\delta n_{j,0} \delta n_{j',t'}}^2 &= \frac{1}{16 \Gamma^2} + {\rm Re}\left(\langle \delta \hat n_{j,0} \delta \hat n_{j',t'} \rangle\right)^2 + \sum_{j_1} {\rm Re}\left(\langle \delta \hat n_{j_1,0} \delta \hat n_{j',t'} \rangle\right)^2.
\end{align}
We used the fact that a zero-mean unit-variance Gaussian random variable $m$ (with $\overline {m} = 0$, $\overline {m^2} = 1$) has the higher order moment $\overline {m^4} = 3$, so $m^2_{j,0} m_{j_1,0}m_{j_2,0} = 2 \delta_{j,j_1}\delta_{j_1,j_2} + \delta_{j_1,j_2}$.
This recovers the scaling of the ``statistical'' noise shown in green in Fig.~3, where the uncertainties are the standard error of the mean, i.e., $[{\rm var} / (M-1)]^{1/2}$, with noise-variance var defined as above.

\subsection{Systematic uncertainties}

In studying the systematic error, we consider fluctuations away from the exact result
\begin{align}
\langle \hat n_{j',t'}\rangle' - \langle \hat n_{j',t'}\rangle &= 2 \Gamma^{1/2} \sum_{j_1} m_{j_1,0} {\rm Re}\left( \langle \delta \hat n_{j_1,0} \delta \hat n_{j',t'} \rangle \right) - \Gamma L_{j',t'},
\end{align}
rather than away from the ensemble average as in the preceding section.
In this case, the cross-correlation function 
\begin{align}
\delta n_{j,0} \delta n_{j',t'} &= \frac{1}{4 \Gamma} m_{j,0} m_{j',t'} + \sum_{j_1}  m_{j,0} m_{j_1,0} {\rm Re}\left( \langle \delta \hat n_{j_1,0} \delta \hat n_{j',t'} \rangle \right) - 2 \Gamma^{1/2} m_{j,0} L_{j',t'}
\end{align}
contain a systematic artifact, which, while absent in the mean, contributes a new term to the noise variance
\begin{align}
{\rm var} &= \frac{1}{16 \Gamma^2} + {\rm Re}\left(\langle \delta \hat n_{j,0} \delta \hat n_{j',t'} \rangle\right)^2 + \sum_{j_1} {\rm Re}\left(\langle \delta \hat n_{j_1,0} \delta \hat n_{j',t'} \rangle\right)^2 + 4 \Gamma L_{j',t'}^2.
\end{align}
This then leads to the observed $\propto \Gamma^{1/2}$ scaling in the systematic uncertainty (purple) in Fig.~3.
This is a hybrid term that mixes systematic and statistical uncertainties, leading to a standard error of the mean that still goes to zero in the large~$M$ limit.
True statistical uncertainties that directly affect the mean therefore require higher-order expressions in the measurement model, which are beyond the scope of this work.

\twocolumngrid

\bibliography{main}

\end{document}